\documentclass[%
 aip,
 amsmath,amssymb,
 reprint,%
]{revtex4-2}

\usepackage{mathptmx}
\usepackage{etoolbox}
\usepackage{color}
\usepackage[svgnames]{xcolor}
\usepackage{amsfonts}
\usepackage{graphicx} 
\usepackage{hyperref} 
\usepackage{cleveref}
\usepackage{enumitem}
\usepackage{tabularx}
\usepackage{dcolumn}

\usepackage{graphicx,epsfig,epstopdf}
\usepackage{amsmath} 
\usepackage{amssymb} 
\usepackage{bm}
\usepackage{color}
\usepackage{subfig}
\usepackage{caption}
\usepackage{bbm}
\usepackage{bbold}
\usepackage{amsmath}
\DeclareMathAlphabet{\mathcal}{OMS}{cmsy}{m}{n}
\newcommand{\be}{\begin{eqnarray}}
\newcommand{\ee}{\end{eqnarray}}
\newcommand{\bc}{\begin{center}}
\newcommand{\ec}{\end{center}}

\newcommand{\nn}{\nonumber}

\usepackage{adjustbox}
\usepackage{xcolor}
\usepackage{mathtools}
\usepackage{soul}
\usepackage{amsmath}
\usepackage{multirow}
\usepackage{relsize}
\usepackage{tabularx}
\usepackage{array}
\usepackage{soul}

\newcommand{\ex}[1]{\langle#1\rangle}
\newcommand{\bigex}[1]{\Big\langle#1\Big\rangle}
\newcommand{\ket}[1]{ |#1\rangle}
\newcommand{\bra}[1]{\langle #1|}

\usepackage{centernot}
\usepackage{cancel}

\begin{document}

\title{Photon entanglement-enhanced multidimensional spectroscopy of exciton correlations in photosynthetic aggregates}

\author{Arunangshu Debnath}
\email{arunangshu.debnath@desy.de}
\affiliation{Center for Free-Electron Laser Science CFEL, Deutsches Elektronen-Synchrotron DESY, Notkestrasse 85, 22607 Hamburg, Germany} 
\author{Shaul Mukamel}
\email{smukamel@uci.edu}
\affiliation{Department of Chemistry and Department of Physics and Astronomy, University of California, Irvine, California 92697, USA}

\date{\today}

\begin{abstract}
\noindent
Nonlinear spectroscopic techniques using entangled photon pairs can provide an opportunity to exploit non-classical correlations encoded in two-photon wavefunctions to manipulate two-exciton wavefunctions. We propose an entangled photon pair-enhanced multidimensional spectroscopic technique that is sensitive to exciton-exciton interactions and correlations at the femtosecond timescale. 
Simulations for a dissipative system, namely, the photosynthetic aggregate reveal the superior ability of entangled photon pairs, compared to both transform-limited and frequency-chirped laser pulses, to manipulate excited-state absorption pathways.
The corresponding spectral features in the two-dimensional spectrogram are interpreted in terms of one- and two-exciton resonances. The signal scales linearly with the incoming intensity of the photon sources.
We show that classifying these resonances using entangled photon source in the perturbative limit allow for probing exciton correlations at the natural energy scale. These insights can be used to explore multi-exciton dynamics in molecular systems using multiphoton entanglement. 
\end{abstract}

\maketitle
\section{Introduction}
In molecular aggregates, spectroscopic characterization of exciton-exciton interactions and correlations yields critical insights into underlying aggregation properties. However, the range of spectroscopic techniques available for exploring their temporal and spectral signatures, particularly in the presence of phonon-induced dissipation, remains limited. These challenges stem, in part, from a lack of techniques specifically tailored to extract correlation induced features hidden in the dynamics of spatially delocalized two-exciton states. Double-quantum coherence (DQC), a nonlinear spectroscopic scheme, was proposed as a viable approach for probing quasiparticle correlation and its dynamical signatures \cite{mukamel2010communications, kim2009two, mukamel2016communication, mukamel2007coherent}.
Among four-wave mixing techniques, photon echo and transient grating primarily investigate ultrafast exciton transport. Although the observed signals may contain contributions from two-exciton states, the associated resonances are often obscured by dominant one-exciton resonances. The double-quantum coherence technique investigates spectral features specifically associated with two-exciton states; the observed signal is sensitive to the differential nature of the two-exciton wavefunction compared to that of the constituent two one-exciton wavefunctions \cite{cundiff2009optical, yang2010probing}.
In previous investigations, double-quantum coherence spectroscopy has aided the exploration of quasiparticle correlations in molecular aggregates \cite{kim2009two, abramavicius2008double}, quantum-confined nanostructures \cite{nardin2014coherent, yang2010probing}, correlated lattice systems \cite{chen2025multidimensional}, atomic gases \cite{dai2012two, gao2016probing, lomsadze2020line}, coupled valence and core-hole excitations in molecules \cite{hua2016study}, vibrations \cite{falvo2009coherent}, exciton and vibrational polaritons \cite{debnath2020entangled, debnath2022entangled, saurabh2016two, debnath2025coherent}.\\
In this theoretical investigation, we deploy the DQC technique to characterize the ultrafast dynamics of molecular exciton aggregates, specifically identifying the resonances that originate from exciton-exciton interactions and correlations. DQC probes the ultrafast dynamics of the two-exciton wavefunctions via interfering contributions of excited-state absorption (ESA) pathways. Signatures of exciton correlations are typically hidden in the two-exciton states, making the technique the most suitable for monitoring and classifying correlation-induced resonances.\\
We perform numerical simulations for the light-harvesting complex, LHCII, which acts as the primary photon energy harvester in photosynthetic aggregates. LHCII hosts a large number of one- and two-exciton states delocalized over sites, each of which undergo ultrafast dephasing due to interaction with phonons. As a result, the spectroscopic identification of correlation-induced resonances become particularly challenging. 
In principle, pulse-shaping techniques are expected to address two-exciton resonances and selectively amplify desired components of the signal. However, given the relatively high density of optical transitions within a narrow spectral range, classical laser sources often face 
challenges stemming from the unavoidable time-frequency bandwidth constraints. The situation becomes even more critical when identical exciton transitions possess comparable dephasing rates.
The entangled photon sources promise to partially alleviate some of these obstacles \cite{bittner2020probing, ishizaki2020probing, fujihashi2021achieving, fujihashi2023probing, fujihashi2024pathway, dorfman2014multidimensional, dorfman2016nonlinear, richter2010ultrafast, debnath2020entangled, debnath2022entangled}, even though identifying a suitable parameter regime where entangled photon pairs outperform their classical counterparts remains a challenging task \cite{landes2021quantifying, raymer2022theory}.\\
\begin{figure*}[ht!]
 \includegraphics[width=.96\textwidth]{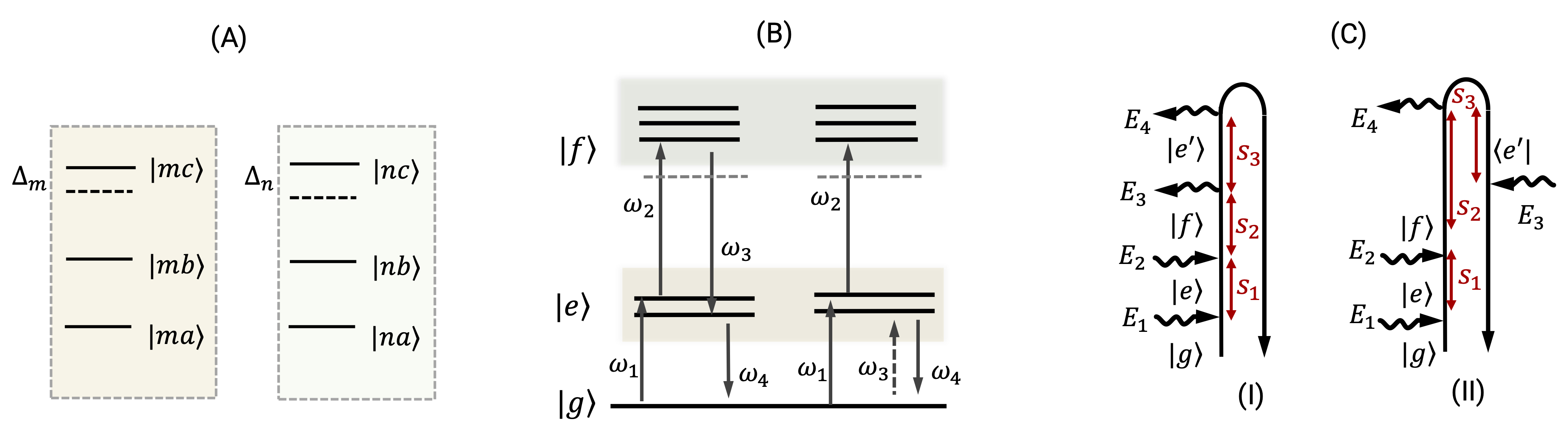}\hfill
  \caption{(A) Schematic representation of two interacting excitonic sites, $m$ and $n$. Each site comprises three energy levels, $\ket{j}$, where $j\in \{a,b,c\})$, highlighting overtone exciton nonlinearities $\Delta_m$ and $\Delta_n$. The model includes 14 such sites. In the delocalized basis, the two-exciton states $\ket{f}$ are affected by both overtone and combination nonlinearities.  (B) Exciton pathways contributing to the signal, shown in Albrecht notation. Processes involving $\omega_1$ and $\omega_2$ are common to both pathways. Processes involving $\omega_3$ and $\omega_4$ determine the differences between the two pathways and are, therefore, discriminative. (C) Corresponding Keldysh-Schwinger loop diagrams for the two pathways contributing to the signal.}
  \label{fig:la01}
\end{figure*}
We demonstrate that the entangled-photon-pair enhanced DQC technique, which utilizes the correlation properties of two-photon wavefunctions, may be suitable for manipulating two-exciton amplitudes. 
It is shown that a superior resolution of two-dimensional spectral features can be obtained compared to both transform-limited and linearly chirped pulses. Further, the spectral features are shown to be dependent on the variation of the exciton nonlinearity parameter, thus identifying their origin in exciton correlations.\\
In what follows, we introduce the exciton-phonon Hamiltonian of the monomeric LHCII, define the one- and two-exciton states, and derive the relevant Green's functions in section~\ref{subsec:ham}. In section~\ref{subsec:result}, which is divided into three subsections, we present the analytical signal and corresponding numerical simulations. Simulations utilize entangled photon sources, classical laser pulses, and a variation of the exciton nonlinearity parameters. The article concludes in section~\ref{sec:conclude} with a summary and an outlook.
\section{Theory}
\subsection{Hamiltonian}\label{subsec:ham}
We construct a description of LHCII monomer, using the Frenkel exciton model within the Heitler-London approximation. For this purpose, we take the lowest-lying adiabatic electronic states of each of the $N_s=14$ chromophoric sites containing chlorophyll molecules.
The combined exciton-phonon Hamiltonian, in the site basis, is given by
\begin{subequations}
\begin{align}
    H_{}^{} 
    &=\sum_{mn} (E_m^{} \delta_{mn} + J_{mn}) B_m^\dag B_n^{}  \\
    &+\sum_{mn} U_{m}^{(1)} B_m^\dag B_m^\dag B_m^{}B_m^{}+U_{mn}^{(2)} B_m^\dag B_n^\dag B_n^{}B_m^{}\\
    &+\sum_{j}^{} \omega_j^{} (b^\dag_j b_j^{}+\frac{1}{2}) + \sum_{m,j}^{} g_{m,j}^{} (b^\dag_j+ b_j^{}) B_m^{\dag} B_m^{}
\end{align} \label{eqn:hamex}   
\end{subequations}
where $B^\dagger_{m}(B_{m}^{})$ represent the exciton creation (annihilation) operators, which follow the commutation relation $ [B_n^{},B_m^\dag] = \delta_{mn}(1- \eta (B_m^\dag B_m^{})^2)$ (where $\eta=3/2$). Parameters $E_m$, $J_{mn}$, $U_{m}^{(1)}$, and $U_{mn}^{(2)}$ represents one-exciton energy, and inter-site hopping, overtone, and combination exciton non-linearity parameters, respectively. $H_{mn}^{(1)} =\sum_{mn} (E_m^{} \delta_{mn} + J_{mn})$ denotes the one-exciton Hamiltonian matrix elements, while $H_{mnkl}^{(2)}= H_{mn}^{(1)} \delta_{kl}^{}+\delta_{mn}^{} H_{kl}^{(1)} + U_{m}^{(1)}+ U_{mn}^{(2)}$ denote their two-exciton counterparts. While, all parameters $J_{mn}$, $U_{m}^{(1)}$ and $U_{mn}^{(2)}$ affect the overtone and combination two-exciton states, the latter two are responsible for the shift in two-exciton energies. It can be expressed as $ E_{m}^{(2)} = 2 E_{m}^{(2)} + \Delta_m$ (where $\Delta_m$ is negative is the two-exciton energy is stabilized). 
In the dipole approximation for the charge densities, $J_{mn}$ are determined by transition dipole moments, while $U_{m}^{(1)}$, $U_{mn}^{(2)}$ are determined by the difference between the permanent dipole moments of the excited and ground states. 
Previous theoretical investigations of two-exciton states based on coupled two-level models have omitted the role of overtone states. Whereas arguments based on laser bandwidth might justify limiting the number of accessible states in the two-exciton manifold, it would amount to neglecting the role of permanent dipole moments. For chlorophyll molecules, the primary pigments in LHCII, the excited-state permanent dipole moment is often similar in magnitude to the transition dipole moment.\\
The Hamiltonian yields $N_{g} = 1$ ground state, $N_{e} = 14$ one-exciton states, and $N_{f} = 105$ two-exciton states; the latter comprises 14 overtone and 91 combination states. The exciton part of the Hamiltonian, given in the first two lines of Eq.~(\ref{eqn:hamex}), is diagonalized to yield exciton-number-conserving manifolds, denoted as $\ket{g}$, $\ket{e}$, and $\ket{f}$. These states are expressed as $\ket{e_j} = \sum_m T^{(1)}_{j, m} B_m^\dag \ket{0}$ and $\ket{f_k} = \sum_{m,n} T^{(2)}_{k, mn} B_m^\dag B_n^\dag \ket{0}$. The transformation matrix elements, $T_{j, m}^{(1)}$ and $T_{k, mn}^{(2)}$, determine the degree of exciton delocalization for a given exciton state.
The exciton states are coupled to collective vibrational modes—the normal modes of intermolecular vibrations—hereafter referred to as phonons. The bare phonon and exciton-phonon coupling Hamiltonians are defined in the third line of Eq.~(\ref{eqn:hamex}). Here, the phonon-mode creation (annihilation) operators $b_j^\dag$ ($b_j$) follow free-boson commutation relations, defined as $[b_j, b_i^\dag] = \delta_{ij}$, while $g_{m,j}$ represents the site-specific coupling strengths.
For LHCII, the phonon spectral density function comprises $N_{b_2}=48$ high-frequency Brownian oscillator modes and $N_{b_1}=1$ low-frequency, overdamped Brownian oscillator mode, expressed as:
\begin{align}
J_{\text{ph}}(\omega)&=2\lambda_0 \frac{\omega\gamma_0}{\omega^2+\gamma_0^2}+\sum_{j=1}^{N_{b_2}} \frac{2\lambda_j \omega_j^2 \omega \gamma_j}{(\omega_j^2-\omega^2)^2+\omega_{}^2\gamma_j^2}    
\end{align}
The associated parameter values are adopted from \cite{novoderezhkin2011intra, novoderezhkin2005excitation, debnath2020entangled, debnath2022entangled}. All simulations were performed at a temperature of $T=273$ K.\\
The exciton-photon interaction Hamiltonian is given by
\begin{align}\label{eqn:hamint}
   & H_{\mathrm{int}}^{}(t) = E V^\dag + E^\dag V =\sum_{j, pq \in \{ge,ef\}} \sqrt{\frac{2\pi \omega_j}{V_j} }\nn\\&
     \Big( a_j^{} \exp{(-i \omega_j t)} 
    \exp{(i \omega_{pq}^{} t)}    d_{pq}^{}\ket{p}\bra{q}
+\mathrm{h.c.} \Big)
\end{align}
The photonic sources probe the excitonic dynamics by inducing transitions between the exciton manifolds; these are captured by the inter-manifold transition operator $V^\dag =d_{pq}^{}\ket{p}\bra{q} $, where $\omega_{pq}^{}$ denotes the exciton energy gap. If the photonic sources can be manipulated with a high degree of control, the inter-manifold exciton transitions and corresponding resonances can be probed in a systematic manner. \\
The Green's functions are defined as $ \mathcal{G}_{pq}(\omega_{}) =i (\omega_{}-z_{pq}^{})^{-1}$, where the exciton resonances associated with inter-manifold transitions are given by $z_{pq}^{}= \omega_{ab} +i \Gamma_{pq}$. The energy gaps, $\omega_{ab}$ and inter-manifold dephasing parameters, $\Gamma_{pq}$, jointly determine the locations of these resonances.
The dephasing parameters are estimated using the line-broadening function, $\Gamma_{pq}^{} =(\gamma_{p}+\gamma_{q'})/2$, where
\begin{align}
    \gamma_{p} (t) &= \sum_{p'} C_{\text{ph}}(\Omega_{pp'}^{}) \sum_m T_{m p}^{} T_{m p}^{} T_{m p'}^{} T_{m p'}^{}
\end{align}
Here, the thermally weighted phonon spectral density, evaluated at the energy gap $\Omega_{pp'}^{}$, is given by 
\begin{align}
    C_{\text{ph}}(\Omega_{}) &= \int_0^{\infty} dt \exp(i\Omega t) \nonumber\\&
    \times\int \frac{d\omega}{2\pi} J(\omega)\Big[\coth{(\beta \omega/2)} \cos{\omega t} \mp i \sin{\omega t}\Big]
\end{align}
The dephasing properties are determined by the resonant phonon modes. 
\section{Results}
\subsection{Theory}\label{subsec:mdcs}
The DQC signal is related to a specific component of the excitonic nonlinear polarization, $k_{\text{s}}=+k_{1}+k_{2}-k_{3}$ generated as a result of four exciton-photon interactions (see Fig.~\ref{fig:la01}). 
Following the derivation outlined in the Appendix~\ref{app:sigderiv}, the final expression is obtained as
\begin{widetext}
\begin{align}\label{eqn:sigdqc2}
     S(\Omega_3, \Omega_2, \Omega_1) &= s_0 \sum_{f,e',e,g} 
   w_{}^{(1)} \frac{\bigex{ E_4^\dag(w_{e'f}) E_3^\dag(w_{e'g}) E_2(w_{fe}) E_1(w_{eg})} }{
 \tilde{F}_{\Omega}^{(1)}\big(\Omega_3,\Omega_2,\Omega_1; z_{e'g}, z_{fg}, z_{eg}\big) }
   + w_{}^{(2)} \frac{\bigex{ E_4^\dag(w_{ge'}) E_3^\dag(w_{e'f}) E_2(w_{fe}) E_1(w_{eg})} }{
      \tilde{F}_{\Omega}^{(2)}\big(\Omega_3,\Omega_2,\Omega_1; z_{fe'}, z_{fg}, z_{eg}\big)}.
\end{align}
\end{widetext} 
The proportionality constant $s_0$ incorporates all coefficients in the expression and serves as a scaling parameter for the numerical simulation. The dipolar spectral weights associated with the two pathways are given by: $w_{}^{(1)} =d_{e'f} d_{e'g} d_{fe} d_{eg}$ and $w_{}^{(2)} = d_{ge'} d_{e'f} d_{fe} d_{eg}$.
We introduced two multidimensional functions 
 \begin{align}
&\tilde{F}_{\Omega}^{}(\Omega_3,\Omega_2,\Omega_1; z_{c}, z_{b}, z_{a})= 
\nonumber\\&
(\Omega_3- z_{c})^{}(\Omega_2-z_{b})^{} (\Omega_1-z_{a})^{}
\end{align}  
The variables $\Omega_j$ are scanned to identify and track excitonic resonances. In a typical two-dimensional spectrum, plotted with $\Omega_2$ and $\Omega_3$ as the axes for a fixed $\Omega_1$, the off-diagonal spectral features primarily represent signatures originating from exciton interactions and correlations. The four-point photon correlation correlation functions in the numerator allow the modulation of respective spectral weights. \\
Spontaneous parametric down-conversion (SPDC) provides a mechanism for the external manipulation of the two-photon wavefunction and the correlation function \cite{pe2005temporal, dayan2005nonlinear}. In the weak down-conversion regime, the photon correlation function is given by: $\ex{E^{\dag}_{}(\omega_4)E^{\dag}_{}(\omega_4)E^{}_{}(\omega_2)E^{}_{}(\omega_1) }= f^{*}(\omega_4^{},\omega_3^{}) f^{}(\omega_2^{},\omega_1^{})$, where we define the joint spectral amplitude (JSA) of the two-photon field
\begin{align}
&f^{}(\omega_{a}^{},\omega_{b}^{})  = \alpha A_p^{}(\omega_{a}^{},\omega_{b}^{})\nonumber\\
& \qquad \times \Big\{ \text{sinc}[\phi(\omega_a, \omega_b)]\times \exp{i\phi(\omega_a, \omega_b)}+a \leftrightarrow b\Big\}   
\end{align}
Here, the transform-limited field profile of the SPDC pump is given by $A_p^{}(\omega_{a}^{},\omega_{b}^{}) = E_0 \sqrt{\pi/\Gamma_{p,0}}\exp{[-(\omega-\omega_{a}-\omega_b)^2/ 4 \Gamma_{p,0}]}$, where $\Gamma_{p,0}$ is the temporal width. The parameter $\alpha$ governs the down-conversion efficiency.
The phase function $\phi(\omega_j, \omega_k) = (\omega_j - \omega_p/2)T_1 + (\omega_k - \omega_p/2)T_2$ is governed by the entanglement time parameter $T_{\text{ent}} = |T_2 - T_1|$.
The latter is crucial for spectroscopic applications. It is determined by the length of the SPDC crystal and the relative group velocities of the photon pairs within the crystal. The square modulus of the JSA, depicted in Fig.~\ref{fig:entcorr}, shows that for strongly anti-correlated pairs, there is a parametric freedom to select two distant frequency values. 
In ultrafast measurements, this flexibility imposes less stringent conditions on the frequencies of the entangled photon pairs. The associated singular values, also presented in Fig.~\ref{fig:entcorr}, are used to estimate the effective correlation between spectral modes \cite{landes2021quantifying, dorfman2016nonlinear, christ2011probing}. 
In comparison, the four-point field correlation function for classical laser pulses factorizes into the product of individual amplitude functions: $\ex{E^{\dagger}_{}(\omega_4^{}) E^{\dagger}_{}(\omega_3^{}) E_{}^{}(\omega_2^{}) E_{}^{}(\omega_1^{})} 
= A^{*}_{p_4}(\omega_4^{}) A^{*}_{p_3}(\omega_3^{}) A_{p_2}^{}(\omega_2^{}) A_{p_1}^{}(\omega_1^{}) $. The signal, therefore, scales quadratically with the intensity of the incident field. 
The frequency domain representation of a transform-limited pulse with a Gaussian profile is given by
\begin{align}\label{eq:pulsetl}
   A_{p_j} (\omega) &= E_0 \sqrt{\pi/\Gamma_{0}}\exp{[-(\omega-\omega_{0,j})^2/ 4 \Gamma_{0,j}]}
\end{align}
where $\Gamma_{0,j}=1/T_{0,j}^{2}=2 \ln 2/\tau_{0,j}^{2}$, $T_{0,j}$ is the temporal half-width at $1/e$, and $\tau_{0,j}$ is the temporal width of the pulse. 
Pulse shaping techniques effectively adds a frequency dependent phase which, restricted to the second-order dispersion (additionally, neglecting the constant phase and the linear order terms), can be expressed as
\begin{align}\label{eq:pulsechirp}
   A_{c, p_j} (\omega) &= E_0 \sqrt{\pi/\Gamma_{0,j}}\exp{[-(\omega-\omega_{0,j})^2/ 4 \Gamma_{j}]}
\end{align} 
The profile above is used for the linearly chirped pulse. Here, $1/\Gamma_j = 1/\Gamma_{0,j} - 2i \phi_{0,j}^{(2)}$, where $\phi_{0,j}^{(2)}$ is related to the linear chirp rate \cite{chatel2003competition, debnath2013dynamics, debnath2013high}. We introduce $\Gamma_j=1/T_{p,j}^{2}+i\alpha_j$, with 
$T_{p,j}=T_{0,j}\sqrt{1+(2\phi_{0,j}^{(2)}/T_{0,j}^{2})^{2}}$ and $\alpha_j=2\phi_{0,j}^{(2)}/[T_{0,j}^{4}+(2\phi_{0,j}^{(2)})^{2}]$.
Here, $\alpha$ is the chirp rate satisfying: $\omega_j(t)=\omega_{0,j}+2\alpha_j t$. Setting the quadratic dispersion to zero, i.e., $\phi_{0,j}^{(2)} = 0$ yields transform-limited laser pulse back. In our case, two classes of optical processes induced by the laser pulses, namely absorption and stimulated emissions, involve a large number of excitonic states. The time-dependent instantaneous frequency is suitable for simultaneously controlling a finite number of transitions within the spectral bandwidth.\\
Conceptually, the generation of the DQC signal can be viewed as a combination of two processes: the initial formation of a state-specific two-exciton coherence, followed by the projection of these coherent oscillations onto one of two possible one-exciton coherence components. 
\begin{figure*}[ht!]
 \includegraphics[width=.96\textwidth]{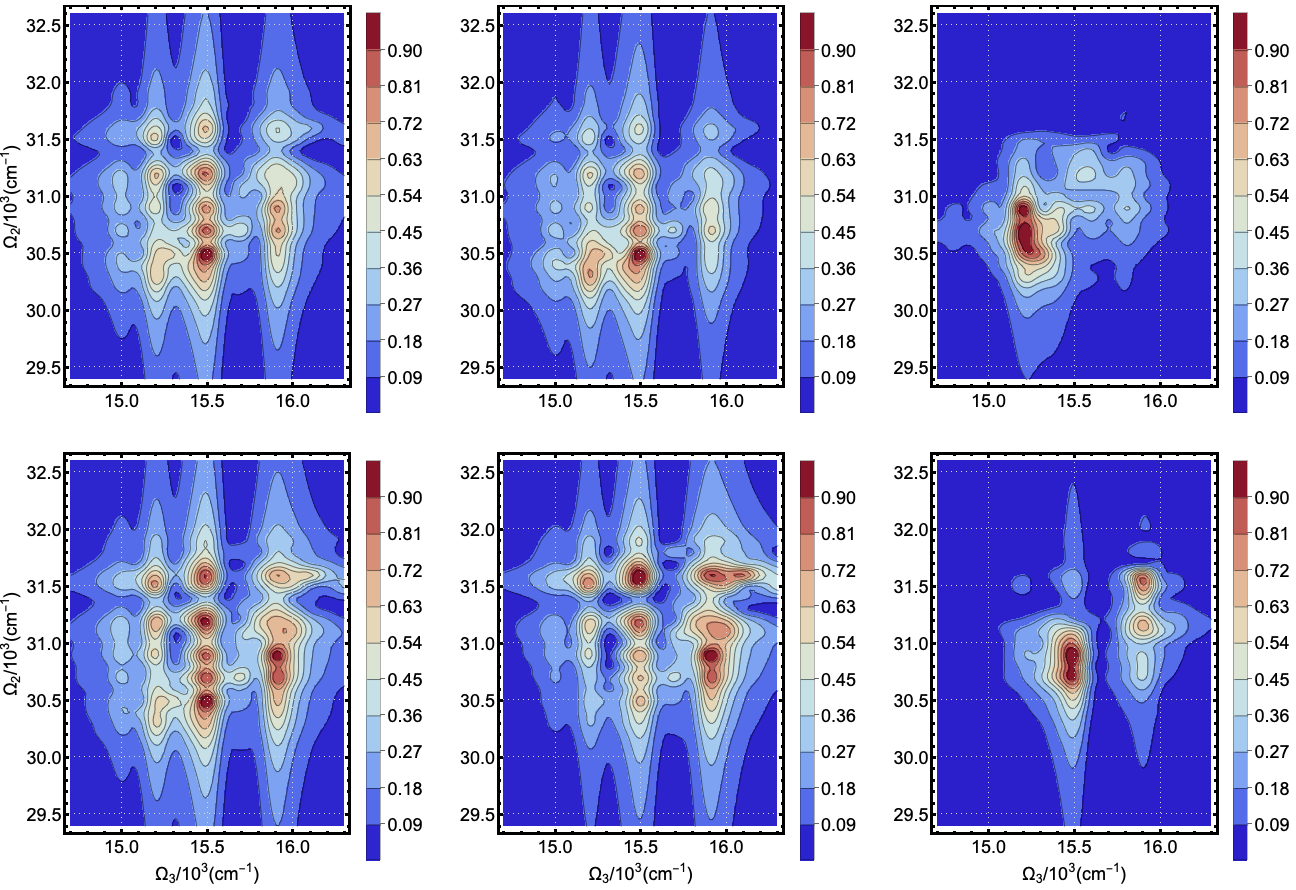}
\caption{Simulated DQC spectra using entangled photon pairs. The sum frequencies of the photon pairs are tuned to the two-exciton states $f_{39}$ (top panel) and $f_{81}$ (bottom panel). The top-left and bottom-left panels serve as reference simulations with $\tilde{T}_{\text{ent}} = 10\, \text{fs}$ and $\tau_0 = 50\,\text{fs}$. The middle and right columns show the effects of varying the temporal width of the SPDC pump and the entanglement time, respectively. The top-middle and bottom-middle panels use $\tau_0 = 100$ fs and $\tau_0 = 20$ fs, respectively, while the top-right and bottom-right panels use $\tilde{T}_{\text{ent}} = 60$ fs. All other parameters remain identical to the reference. For a detailed discussion, see Section \ref{subsec:resultQ}.}
  \label{fig:dqc3981}
\end{figure*}
The two pathways depicted in Fig.~\ref{fig:la01} captures this phenomenology. Here, the two-exciton coherence is generated during the interval $s_1+s_2$, and the one-exciton coherence components are generated during the final interval $s_3$. 
We can manipulate the two-exciton resonances, $z_{fg}$, by modulating the sum frequency $\omega_p = \omega_1 + \omega_2$ during the initial pair of exciton-photon interactions. 
We can also manipulate the one-exciton resonances at $z_{e'g}$ or $z_{fe'}$, by modulating $\omega_3$ and $\omega_4$, during the subsequent pair of exciton-photon interactions. The final signal is the resultant of two interfering pathways, the magnitude of which is largely determined by the last two interactions. 
If the contributions from exciton nonlinearities are not significant, the resonances $z_{e'g}$ and $z_{fe'}$ yield nearly identical contributions. In such situations, we have $z_{fe'} \approx \omega_{e'g} + \Delta + i\Gamma_{fe'} \approx \omega_{e'g} + i\Gamma_{fe'}$, where the energetic deviation $\Delta$ is negligible. Consequently, the two pathways interfere destructively, and the spectral weights diminish toward zero.
\begin{figure*}[ht!]
 \includegraphics[width=.96\textwidth]{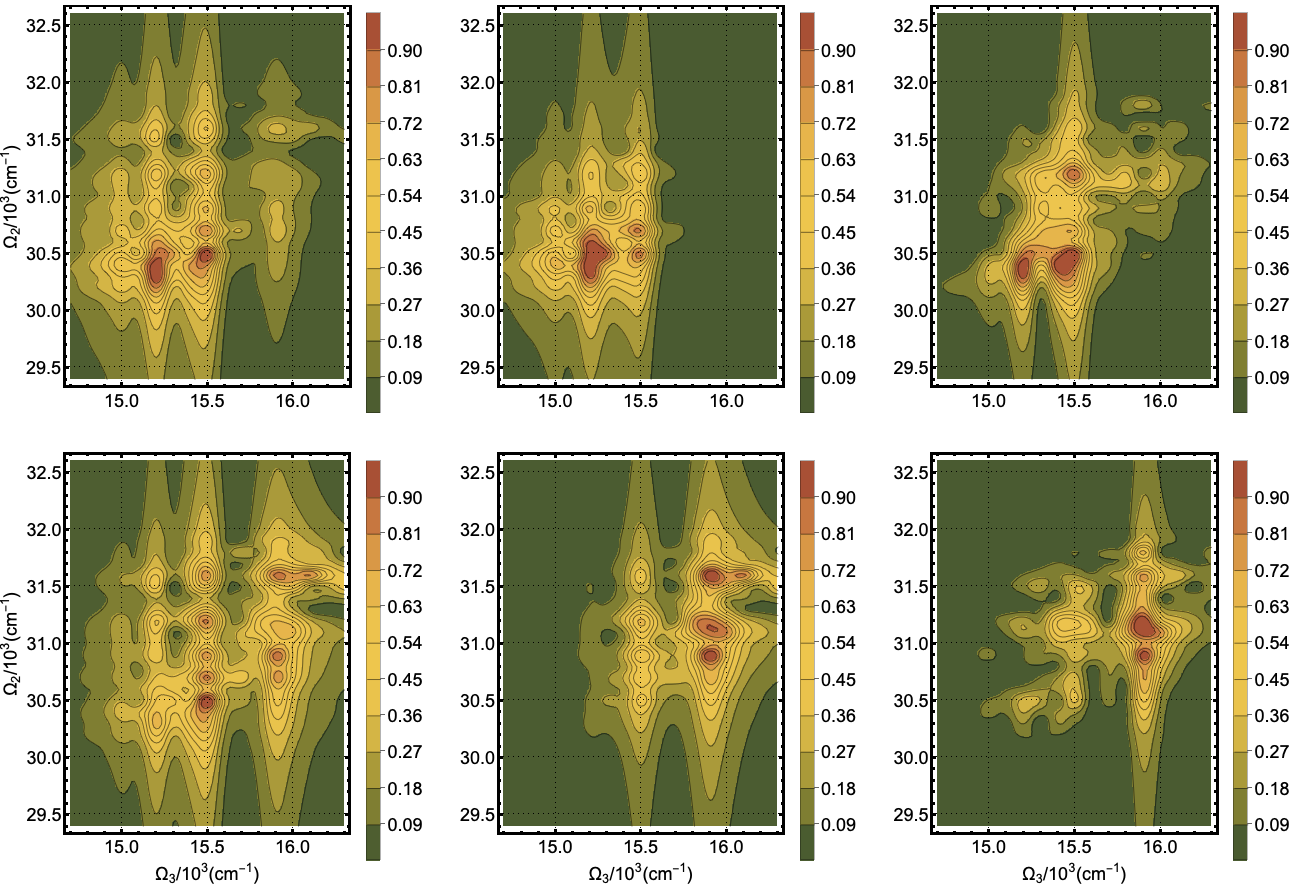}
\caption{Simulated DQC spectra using classical laser pulses. The central frequencies of the laser pulses match the frequencies of the entangled photon pairs deployed in Fig. \ref{fig:dqc3981}. The top- and bottom-left column present the reference simulations with $\tau_{j,0} = 10$ fs. The middle column shows the effect of an increased temporal width by using $\tau_{j,0} = 20$ fs. The decrease in the spectral bandwidth suppresses several peaks. The right column displays spectra obtained using linearly chirped pulses. The negatively chirped pulse was unable to recover the suppressed peaks. For a detailed discussion, see Section \ref{subsec:resultC}.}
  \label{fig:dqc3981c}
\end{figure*}
\begin{figure*}[ht!]
\includegraphics[width=.96\textwidth]{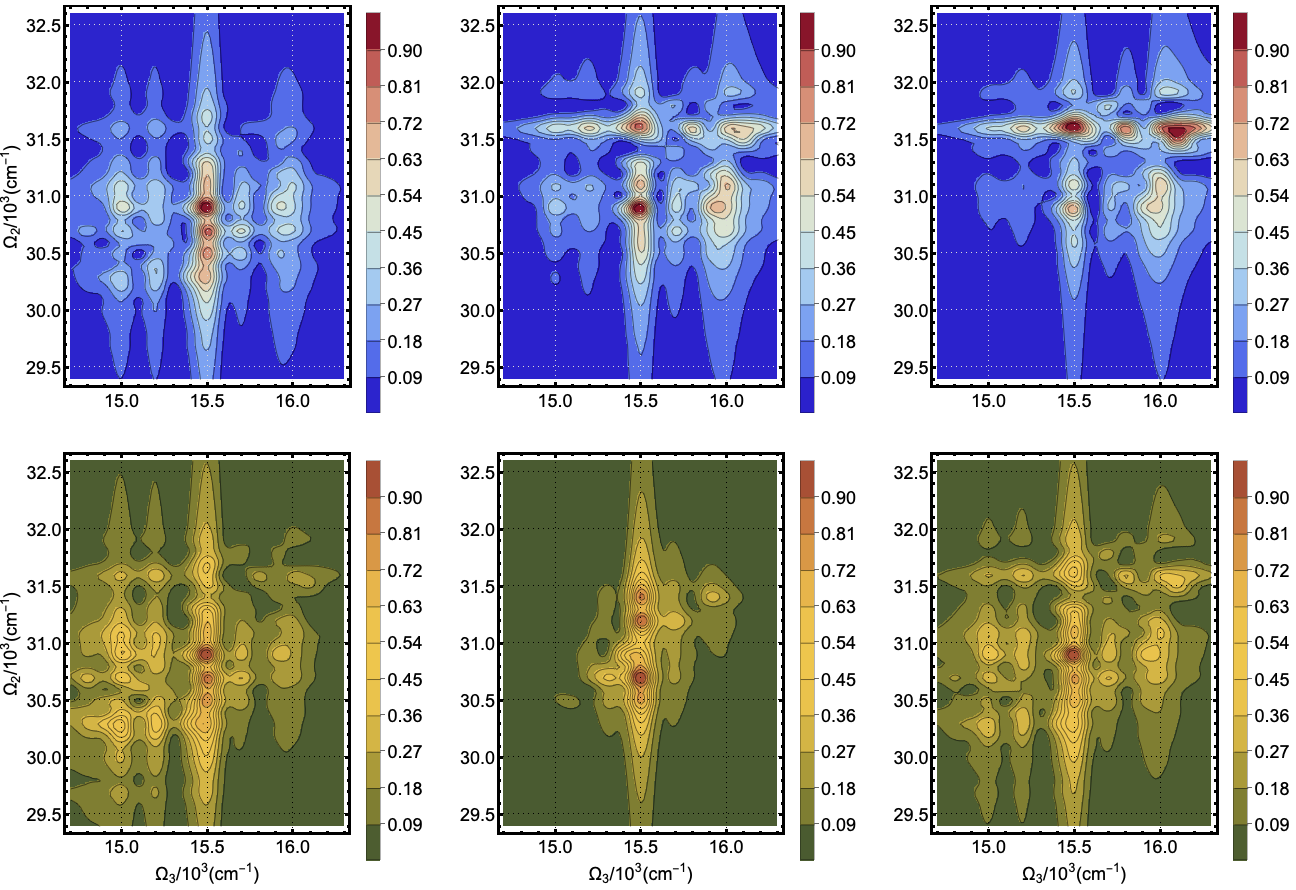}
  \caption{Simulated DQC spectra evaluated using two distinct sets of excitonic nonlinearity parameters. These parameters influence both the two-exciton wavefunction and the transition dipole moments, as highlighted in the Appendix~\ref{app:hamjandk}. The top and bottom panels present simulations using the entangled photon pairs and classical laser pulses, respectively. The sum frequencies are tuned to the two-exciton states $f_{39}$ and $f_{81}$ and we use $\tilde{T}_{\text{ent}} = 10\, \text{fs}$ and $\tau_0 = 50\,\text{fs}$.
  The upper panels utilize larger overtone excitonic nonlinearities (values given in the lower panel of the Fig.~\ref{fig:umag}) than previous simulations in Fig.~\ref{fig:dqc3981}. The top-left and top-middle panels employ the same photonic parameters as those in the top-left and bottom-left of the same. The top-right panel uses degenerate entangled photon pairs with frequencies $\omega_j= E_{e_{13}}$. The bottom-left and bottom-middle panels use larger and smaller overtone excitonic nonlinearities, respectively. The bottom-right panel employs degenerate laser pulses with frequencies $\omega_j= E_{e_{09}}$ to explore the down-shifted resonances. For a detailed discussion, see Section~\ref{subsec:resultU}.}
  \label{fig:sigdqcu}
\end{figure*}
\subsection{Simulations} \label{subsec:result}
Strategies for observing desired spectroscopic features can be devised by jointly considering the exciton gaps $\omega_{pp'}$ and the dephasing parameters $\Gamma_{pp'}$. They jointly contribute to the inter-manifold exciton resonances: $z_{pp,}$, defined earlier.
In an ideal scenario, scanning the $\Omega_2$ axis would maximally display $N(N-1)/2+N =105$ number of $z_{fg}$ resonances, while scanning the $\Omega_3$ axis would display a mixture of $z_{fe}$ and $z_{eg}$ resonances, amounting to  $(N(N-1)/2+N) N =105\times 14$, and $N=14$, resonances respectively. In general, a high density of optical transitions makes it difficult to control individual spectral features independently. Additionally, the presence of dephasing causes several resonances to overlap. The external photonic sources may also mask several resonances depending on its spectral bandwidth properties.
While intuition suggests that a narrowband source is a reasonable choice for generating state-specific two-exciton coherence, it may face challenges by the ultrafast dephasing of the mediating one-exciton states. Similarly, the selective projection to one-exciton coherence using a narrowband source can be less discriminating if the state transitions involved have similar spectral weights.
These problems can be partially mitigated by exploiting the properties of the entangled photon pairs.\\
In the following, we present numerical simulations across three subsections. First, we deploy entangled photon pairs to explore correlations encoded in a two-exciton state by exciting the system via a selected pair of one-exciton states. Two such cases are explored. For each of these cases we choose a different pair of mediating one-exciton states.
Second, we use a transform-limited pulse and a frequency-chirped pulse to recover the spectral features observed in the first case. A principal aim of this study is to recover the maximum number of exciton resonances observed in the previous case by utilizing both broadband and narrowband classical photonic sources.
Third, we repeat some of the simulations for two different exciton nonlinearity parameters and degenerate laser pulses. The exciton nonlinearity parameters quantify exciton correlation within the Frenkel exciton model. This study aims to explore the alteration of the previously observed spectral features. 
\subsubsection{DQC spectra with entangled photon pairs}\label{subsec:resultQ}
In Fig.~\ref{fig:dqc3981}, we present the simulated two-dimensional spectra obtained using entangled photon pairs. Two distinct two-exciton states, $f_{39}$ and $f_{81}$, are targeted by tuning the sum frequency $\omega_p$ to the respective state energies (presented in the top and bottom panels, respectively). Furthermore, the frequencies of the entangled photon pairs are selected such that $\omega_1=\omega_3$ and $\omega_2=\omega_4$, implying that the first pair of photons is identical to the second.
We also apply the following energy matching conditions for selecting the mediating one-exciton states: $\omega_1 \approx E_{e_{05}}$ and $\omega_2 \approx E_{e_{09}}$ for the top panel (where $\omega_p \approx E_{f_{39}}$), and $\omega_1 \approx E_{e_{07}}$ and $\omega_2 \approx E_{e_{13}}$ for the bottom panel (where $\omega_p \approx E_{f_{81}}$). 
The corresponding state energies are $E_{f_{39}} = 30765.4 \,\text{cm}^{-1}$, $E_{f_{81}} = 31219.6 \,\text{cm}^{-1}$ for the targeted states, and $E_{e_{05}} = 15221.6 \,\text{cm}^{-1}$, $E_{e_{07}} = 15370.1 \,\text{cm}^{-1}$, $E_{e_{09}} = 15655.9 \,\text{cm}^{-1}$, $E_{e_{13}} = 15907.2 \,\text{cm}^{-1}$ for the mediating states. \\
The results in the top-left and bottom-left panels utilize $\tau_p=50\,\text{fs}$, $T_{\text{ent}}= 10\,\text{fs}$, which serve as the reference simulation. In the middle and right columns, we vary the temporal width of the SPDC pump, $\tau_p$, and the entanglement time, $T_{\text{ent}}$, respectively.
For the top-middle and bottom-middle panels, we use $\tau_p=100\,\text{fs}$ and $\tau_p=20\,\text{fs}$, respectively. 
For both panels in the right column, we set $T_{\text{ent}}=60$ fs. 
\\
The top-left panel shows a series of peaks which can be partitioned into five sectors along the $\tilde{\Omega}_3$ axis and four sectors along the $\tilde{\Omega}_2$ axis (where $\tilde{\Omega} = \Omega/10^3$ in $\text{cm}^{-1}$).
Along the $\tilde{\Omega}_2$ axis, these spectral features can be identified can be identified as $31.5$ ($\approx z_{f_{51} g}$), $30.9$ ($\approx z_{f_{49} g}$), $30.4\text{--}30.5$ ($\approx z_{f_{25} g}\text{--}z_{f_{27} g}$), and $30.2$ ($\approx z_{f_{19} g}$). Identifying the peaks along the $\tilde{\Omega}_3$ axis is more cumbersome due to interfering contributions. However, we approximately identify the peaks with the following inter-manifold resonances: $15.0$ ($\approx z_{f_{38} e_{9}}\text{--}z_{f_{38} e_{10}}$), $15.3$ ($\approx z_{f_{38} e_{8}}, z_{f_{39} e_{9}}$), $15.5$ ($\approx z_{f_{38} e_{3}}, z_{f_{39} e_{3}}, z_{f_{40} e_{3}}$), and $15.9\text{--}16.0$ ($\approx z_{f_{41} e_{4}}$).
In the top-middle panel, changing the temporal width produced several noticeable changes; the spectral weights of several off-diagonal peaks are drastically reduced. The reduction is attributed to lowering of the spectral bandwidth which allows fewer pathways.
Increasing the entanglement time alters the intervals between successive exciton-photon interactions. In the top-right panel, it leads to the vanishing of off-diagonal peaks. Only the peaks at $(\tilde{\Omega}_2, \tilde{\Omega}_3) \approx (30.9, 15.3)$ survives in the top-right panel, while $(\tilde{\Omega}_2, \tilde{\Omega}_3) \approx (31.3, 16.0)$ survives in the bottom-right panel.\\
The bottom-left panel exhibits peaks similar to those in the top-left panel, except for the off-diagonal peaks at $(\tilde{\Omega}_2, \tilde{\Omega}_3) \approx (31.5, 15.5)$ and within the range $(\tilde{\Omega}_2, \tilde{\Omega}_3) \approx (31.6\text{--}31.9, 16.0)$ which are now pronounced.
An interesting case emerges when increasing the temporal width of the SPDC pump: several off-diagonal peaks survive at $(\tilde{\Omega}_2, \tilde{\Omega}_3) \approx (31.5, 15.3)$, $(31.5, 15.5)$, $(30.7, 15.5)$, and $(30.7, 16.0)$. The larger spectral bandwidth allows some new peaks while retaining most of the earlier ones. This can be explained by destructive interference among the pathways allowed by the larger bandwidth.
The bottom-right panel demonstrates a marked loss in the number of spectral features, with the exception of a surviving off-diagonal peak at $(\tilde{\Omega}_2, \tilde{\Omega}_3) \approx (31.3, 16.0)$.
Overall, we note that the spectra in both panels retain their qualitative features along the $\tilde{\Omega}_3$ axis, even though they exhibit differences along the $\tilde{\Omega}_2$ axis. In contrast to the top panel, the bottom panel exhibits reduced sensitivity to changes in the temporal width of the SPDC pump. This observation highlights that exciton correlation contributions manifest differently for the targeted $\ket{f}$ states.
\subsubsection{DQC spectra with transformed limited and chirped laser pulses}\label{subsec:resultC}
In Fig.~\ref{fig:dqc3981c}, we present the two-dimensional spectra simulated using two sets of classical laser pulses: transform-limited and chirped laser, both characterized by a Gaussian profile. The central frequencies correspond to those utilized for the entangled photon pairs, meaning that the targeted and mediating states remain consistent with those in Fig.~\ref{fig:dqc3981}. For the reference simulations, presented in the left column of both panels, we employ a transform-limited pulse with a temporal width of $\tau_{j,0} = 10\,\, \text{fs}$ .
We explore the spectral features resulting from the variation of the temporal width, by setting $\tau_{j,0} = 20$ fs, and the application of an additional spectral phase, by using a linear chirp parameter of $-750$ fs$^2$, in the middle and right columns, respectively. \\
In the left column, we observe that compared to the previous simulations in Fig.~\ref{fig:dqc3981}, the number of spectrally resolved resonances is reduced, particularly along the $\Omega_3$ axis. Several spectral peaks exhibit marked broadening, and many carry diminished spectral weights. These features persist across all the plots. We also note that the top-left panel exhibits a loss of more number of off-diagonal peaks compared with the bottom-left panel.
In particular, the spectra in the top-left exhibit dominant off-diagonal peaks at $(\tilde{\Omega}_2; \tilde{\Omega}_3) \rightarrow (30.5; 15.5), (30.7; 15.5)$. The longer temporal width of the pulse indicate relatively narrow spectral width. The middle panel reveals that compared to the left panels, the majority of the peaks in the top- and bottom-middle panels are absent at higher and lower values of $\tilde{\Omega}_3$, respectively. In the bottom-middle panel, two peaks at $(\tilde{\Omega}_2; \tilde{\Omega}_3) \rightarrow (30.7; 16.0), (30.2; 16.0)$ gained spectral weights.
In negatively chirped pulses, the instantaneous frequency decreases over time. In other words, the high-frequency components arrive earlier within the pulse profile. Such pulses can be used to influence the discriminative stage of the DQC signal, where stimulated emission involving one-exciton resonances occurs. The limited capacity of the chirp parameters to resolve multiple resonances is reflected in the spectra. We observe a few peaks at $(\tilde{\Omega}_2; \tilde{\Omega}_3) \rightarrow (30.4; 15.5)$ in the top-right panel and $(\tilde{\Omega}_2; \tilde{\Omega}_3) \rightarrow (31.2; 16.0)$ in the bottom-right panel gaining significant spectral weights while others vanish. 
In drawing this conclusion, we note that since the classical laser frequencies address the same mediating one-exciton states, the specific combination of the chirp parameters and central frequencies is unfavorable. This result also highlights the dominant role played by two-exciton coherence generation stage. 
\subsubsection{DQC spectra for a variation of the exciton nonlinearity parameters and degenerate frequencies}\label{subsec:resultU}
As described by the second term of Eq.~\ref{eqn:sigdqc2}, the spectral peaks and their respective weights are dictated by interference between two pathways. These pathways contain inter-manifold resonances that correspond to the energy gaps $\omega_{fe'}$ and $\omega_{e'g}$; the differences between these resonances determine the overall magnitude of the signal. In the Frenkel exciton description, these differences are largely due to the exciton nonlinearity parameters $U_{m}^{(1)}$ and $U_{mn}^{(2)}$. The exciton nonlinearities, while directly affecting the two-exciton eigenenergies, also changes the transformation matrices and, consequently, the transition dipole moments.
Figure~\ref{fig:sigdqcu} illustrates the influence of the variation of exciton nonlinearity parameter $U_{m}^{(1)}$ on the spectra, presented in the left and middle columns. Note that $U_{m}^{(1)}$ values were negative in previous sections, signifying the energetic stabilization of overtone two-exciton energies; the distribution of these values is shown in Fig.~\ref{fig:umag}.
We also investigate the effect of degenerate entangled photon pairs and degenerate classical lasers on the spectra, presented in the top- and bottom-right panels, respectively.\\
The top-left and top-middle panels repeat the reference simulation from Fig.~\ref{fig:dqc3981}, but the nonlinearity parameters $U_{m}^{(1)}$ are set to higher values, respectively. 
Specifically, these parameters are increased to the values depicted in Fig.~\ref{fig:umag}. 
Setting the magnitude of the $U_{m}^{(1)}$ higher effectively shifts the energy gaps. This change dominantly affects two-exciton eigenstates that have a high overlap with the overtone states in the site basis.
We note that in the top-left panel, compared with the reference in Fig.~\ref{fig:dqc3981}, the peaks along $\tilde{\Omega}_3 \approx (15.0, 15.3, 16.0)$ are the most affected. In general, their spectral weights are diminished, and the peaks are clustered within the range $\tilde{\Omega}_2 \approx (30.2\text{--}31.0)$.\\
In a similar spirit, in the lower panels, we use classical transform-limited laser pulses with parameters identical to those of the reference simulation in Fig.~\ref{fig:dqc3981c}. 
The nonlinearity parameters are set to higher and lower values in the bottom-left and bottom-middle panels, respectively. The latter makes these values in the bottom-middle panel negligibly small.
We note that in the bottom-left panel, the peaks around $(\tilde{\Omega}_2, \tilde{\Omega}_3) \approx (30.3, 15.5)$ exhibit reduced spectral weight, following a trend similar to that observed for the entangled photon pairs. The previous argument remains valid: the shift in the resonances rendered the existing laser parameters unable to track the resonance frequencies.
In the bottom-middle panel, due to the reduction of the exciton nonlinearity parameter by one order of magnitude, we approach a limit where two-exciton eigenenergies are approximately equal to the sum of one-exciton energies. This implies that the respective $z_{fe'}$ resonances are of approximately equal strength to the $z_{e'g}$. Consequently, the spectra are devoid of most correlation features, except for minor residuals.\\
The top-right panel utilizes degenerate entangled photon pairs to revisit the scenario presented in the top-middle panel. These frequencies are selected so that the mediating one-exciton states become $\ket{e_{13}}$. We note that two off-diagonal zones around $(\tilde{\Omega}_2, \tilde{\Omega}_3) \approx (31.5, 15.5)$ and $(\tilde{\Omega}_2, \tilde{\Omega}_3) \approx (31.5, 16.0)$, become prominent, while the diagonal peaks at $(\tilde{\Omega}_2, \tilde{\Omega}_3) \approx (30.7, 15.5)$ are diminished.
The bottom-right panel utilizes degenerate classical laser pulses to revisit the scenario from the bottom-left panel. Here, we choose the central frequencies to ensure that the mediating one-exciton states remain $\ket{e_{09}}$. All residual off-diagonal resonances are now markedly diminished. \\
\section{Conclusion and outlook}\label{sec:conclude}
In this article, we demonstrated that the non-classical correlation of entangled photon pairs can be utilized to enhance the resolution of two-dimensional double quantum coherence spectra, even in the presence of dephasing. The proposed scheme can be realized by combining phase-cycling techniques with interferometric detection  \cite{raymer2013entangled, dorfman2014multidimensional, debnath2020entangled}. 
The mathematical expressions in Eq.~\ref{eqn:sigdqc2} in section~\ref{subsec:mdcs} reveal that the nonlinear excitonic response involves three types of inter-manifold resonances: $z_{fg}$, $z_{fe}$, and $z_{eg}$. The entangled photon pairs were able to monitor and selectively amplify some of these resonances in the spectra, as shown in Fig.~\ref{fig:dqc3981}.
Molecular aggregates which are composed of multiple sites, each hosting multiple energy levels, the number of plausible optical transitions are often large. It leads to a large number of dynamical resonances; they may appear congested in the spectra. Furthermore, when weakly coupled phonon modes act as reservoirs, two or more dynamical resonances may involve distinct transition energies yet possess similar dephasing rates; they appear coalesced in the spectra.
A well characterized source of entangled photon pairs can simultaneously control the delays between successive exciton-photon interactions and the combination frequencies of the photons. This, in turn, may enable the discrimination between the dynamical resonances, leading to the enhancement of the spectral resolution across either of the parametric axes, $\tilde{\Omega}_2$ or $\tilde{\Omega}_3$.
Classical transform-limited pulses, while otherwise suitable for probing ultrafast dynamics, have proved to be inadequate in the investigated scenarios. As shown in Fig.~\ref{fig:dqc3981c} and discussed in Section \ref{subsec:resultC}, the spectral weights associated with the two pathways could not be modulated over the full parameter regime, resulting in only a partial suppression or amplification of the resonances.
In certain cases, appropriate spectral phase modulation is known to circumvent some of the previous issues by spreading the frequency components across the pulse profile. This led us to explore the role of frequency-chirped laser pulses, discussed in the Section~\ref{subsec:resultC}. We note that even though a more sophisticated pulse-shaping protocols can be used \cite{debnath2012chirped, abramavicius2008double, hedse2023pulse, lavoie2020phase, debnath2013high, fersch2023single}, the signal would still exhibit a quartic scaling with the intensity of the incoming field.\\ 
The exciton nonlinear response and the resulting two-dimensional spectra were computed using a microscopic model of the photosynthetic complex, LHCII. The simulation indicates that a complex interplay of exciton hopping, dephasing, and exciton nonlinearity governs the spectral weights of the signal. In previous studies, ultrafast energy transfer in the one-exciton manifold was spectroscopically investigated \cite{do2019revealing, maly2016single, zhang2015direct, arsenault2020role}. These studies often omit the role of excited-state absorption pathways, which are essential for the generation of the DQC signal. The simulation uses a set of realistic exciton and phonon parameters at finite temperature, parameters of which were taken from \cite{novoderezhkin2011intra, debnath2022entangled, debnath2026entangled}. \\
Within the Frenkel exciton model, the exciton nonlinearity parameter captures the effect of exciton correlations, the latter originating from Coulomb interactions. 
The simulation accounts for both one- and two-exciton states; the two-exciton Hamiltonian is explicitly dependent on exciton nonlinearity parameters. The inter-manifold resonances, via the diagonalization step, carry this dependency into the expression of the signal. To examine the correlation origin of some of the observed features, we varied the exciton nonlinearity parameter and studied the resulting spectral modulation. It is discussed in Section~\ref{subsec:resultU}. 
It is demonstrated that lowering the nonlinearity parameter caused most resonances to vanish, while increasing it gave rise to a new set of resonances. The strategy bolsters the capability of the DQC signal to track spectral features arising from exciton interactions and correlations across a diverse parameter regime. We note that due to the presence of multiple levels in two different manifolds, a complete suppression of the signal can not be expected. \\
The Coulomb interactions involving two-exciton transitions also lead to several interesting phenomena, such as exciton fusion, scattering, and annihilation. These processes await a detailed spectroscopic investigation in the ultrafast regime. In our simulations, we strategically focus on the two-exciton states $f_{39}$ and $f_{81}$, which lie in the lower and upper halves of the two-exciton manifold and may be prone to exciton-exciton annihilation (EEA).
Even though the spectral signatures of EEA are typically monitored via fifth-order nonlinear optical signals, lower-order signals, such as DQC, can serve as an early-stage characterization tool. The DQC technique also provides a protocol for assigning state-specific dephasing parameters for the involved two-exciton states. The parameter regime explored in this work provides a transparent guide for measurements seeking to quantify excited-state dephasing parameters using entangled photons \cite{liu2022probing}. These measurements would be useful for both spectroscopy and microscopy \cite{tsao2025enhancing, casacio2021quantum}.\\
The parameters of the entangled photon pairs are determined by the properties of the Joint Spectral Amplitude (JSA), depicted in Fig.~\ref{fig:entcorr} in Appendix~\ref{app:photonjsa} and discussed in Section~\ref{subsec:mdcs}. A spectral engineering approach to the JSA would offer a consistent framework for determining the SPDC parameters \cite{dosseva2014shaping, shukhin2024two, weiss2025nonlinear}.
The principal outcome of this work is the ability to explore and monitor the survival of spectral signatures originating from interfering pathways. The pathways in the signal expression appear in a convolutional form containing exciton and photon correlation functions; it offers ample room for adopting open-loop control techniques. 
Control objectives may seek to optimize the photon correlation functions such that specific terms in the sum-over-states expression are completely suppressed. It will provide an extension to the approach adopted in this work, where the entangled photon parameters were microscopically informed yet intuitively estimated. 
We note that applying the proposed technique to monitor the dynamics in correlated materials at the Mott limit provides an exciting opportunity \cite{citeroni2025ultrafast, debnath2023theory, nambiar2025diagnosing, chen2025multidimensional}, which will be addressed in a subsequent communication. 
\begin{acknowledgments}
A.D. acknowledges the support from DESY (Hamburg, Germany), a member of the Helmholtz Association HGF. S.M. gratefully acknowledges the support of the National Science Foundation (NSF) grant Grant No. CHE-2246379.
\end{acknowledgments}
\appendix
\section{Notes on the Hamiltonian}\label{app:hamjandk}
\begin{figure}[ht!]
\includegraphics[width=.48\textwidth]{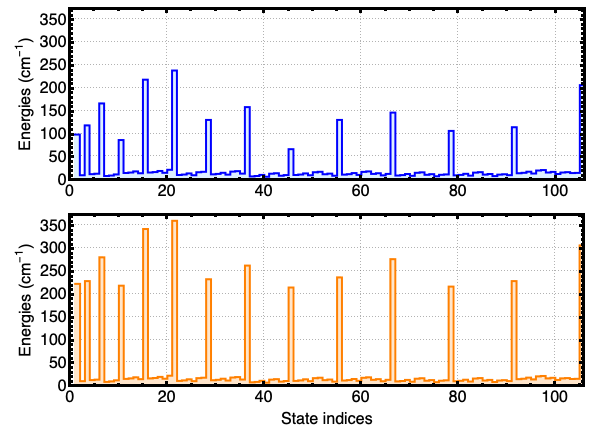}
  \caption{The absolute values of the exciton nonlinearity parameters are plotted. The sequence follows the order in which the values appear in the local two-exciton basis. Values in the upper panel are used in the simulations for Fig.~\ref{fig:dqc3981} and Fig.~\ref{fig:dqc3981c}, while those in the lower panel are used in Fig.~\ref{fig:sigdqcu}. Further details regarding the Hamiltonian and these results are provided in Section~\ref{subsec:ham} and ~\ref{subsec:resultU}.}
 \label{fig:umag}
\end{figure}
The exciton Hamiltonian contains two types of exciton nonlinearity parameters, namely $U_{m}^{(1)}$ and $U_{mn}^{(2)}$. They affect overtone and combination two-exciton terms in the Hamiltonian, via $H_{mnkl}^{(2)}= H_{mn}^{(1)} \delta_{kl} + \delta_{mn} H_{kl}^{(1)} + U_{m}^{(1)}+U_{mn}^{(2)}$. In this work, we use two sets of exciton nonlinearity parameters that differ by the magnitudes of the overtone nonlinearities only. The absolute magnitudes of these two sets are plotted in Fig.~\ref{fig:umag}.
\section{Outline of derivation of the signal}\label{app:sigderiv}
In this appendix, we succinctly outline the derivation, directing the reader to \cite{debnath2022entangled, richter2010ultrafast, dorfman2016nonlinear} for a more involved discussion. 
The DQC signal is defined as the integrated rate of change of the transmitted intensity of the detected photon mode, denoted $E_{4}$. It is expressed as $S = \int_{-\infty}^{\infty} dt_4 \langle E_{4}^\dag (t_4) E_{4} \rangle$. Here, the averaging is performed over the time-dependent joint exciton-photon density operator which involves partial trace over the phonon modes. Expanding the density operator to the third power of the exciton-photon interaction, and introducing the the frequency domain variables, we obtain:
\begin{widetext}
\begin{align}\label{eqn:sigdqc1}
    S &=\int_{-\infty}^\infty dT_1 dT_2 dT_3 \,\, F_{}(\Omega_3, \Omega_2, \Omega_1; T_3, T_2, T_1) 
     \int_{-\infty}^{\infty} \frac{d\tilde{\omega}_3}{2\pi}\frac{d\tilde{\omega}_2}{2\pi}\frac{d\tilde{\omega}_1}{2\pi} e^{ -i (-\tilde{\omega}_3+\tilde{\omega}_2+\tilde{\omega}_1)T_3 -i (\tilde{\omega}_1+\tilde{\omega}_2) T_2-i\tilde{\omega}_1 T_1}\nonumber\\
    &  \Big\{  \ex{E_3^\dag(\tilde{\omega}_3) E_4^\dag(+\tilde{\omega}_1+\tilde{\omega}_2-\tilde{\omega}_3) E_2(\tilde{\omega}_2) E_1(\tilde{\omega}_1)} 
 \sum_{\substack{f,e',e,g}}
     w_{}^{(1)}
    \mathcal{G}^{}_{e'g}(\tilde{\omega}_1+\tilde{\omega}_2-\tilde{\omega}_3) \mathcal{G}_{fg}^{}(\tilde{\omega}_2+\tilde{\omega}_1) \mathcal{G}_{eg}^{}(\tilde{\omega}_1)\nonumber\\
    &+
    \ex{E_3^\dag(\tilde{\omega}_3) E_4^\dag(+\tilde{\omega}_1+\tilde{\omega}_2-\tilde{\omega}_3) E_2(\tilde{\omega}_2) E_1(\tilde{\omega}_1)} 
    \sum_{\substack{f,e',e,g}}
    w_{}^{(2)}
    \mathcal{G}_{fe'}^{}(-\tilde{\omega}_3+\tilde{\omega}_2+\tilde{\omega}_1) 
    \mathcal{G}_{fg}^{}(\tilde{\omega}_2+\tilde{\omega}_1) \mathcal{G}_{eg}^{}(\tilde{\omega}_1) \Big\}.
\end{align}
\end{widetext}
In Eqn. \ref{eqn:sigdqc1}, we introduced the integral transform operator,
\begin{align}
    & F_{}(\Omega_3, \Omega_2, \Omega_1; T_3, T_2, T_1) = \prod_{j=1}^{3} \theta(T_j) \exp({i\Omega_j T_{j}} )
\end{align}
where the delay variables, $T_p$, represent the differences between the centering times of successive photon-matter interactions. Specifically, these delays are defined as: $\tilde{\tau}^0_{21}\rightarrow T_1, \tilde{\tau}^0_{32}\rightarrow T_2, \tilde{\tau}^0_{43}\rightarrow T_3$, for the diagram I, and $\tau^0_{21} \rightarrow T_1, \tau^0_{42}-\tau^0_{43} \rightarrow T_2, \tau^0_{43}\rightarrow T_3$, for the diagram II. 
Starting from Eq. (\ref{eqn:sigdqc1}), we perform the frequency integrations to obtain the final expression for the signal, presented in Eq.~\ref{eqn:sigdqc2}.
\section{Notes on the Joint spectral intensity function}\label{app:photonjsa}
\begin{figure*}[ht!]
\includegraphics[width=.96\textwidth]{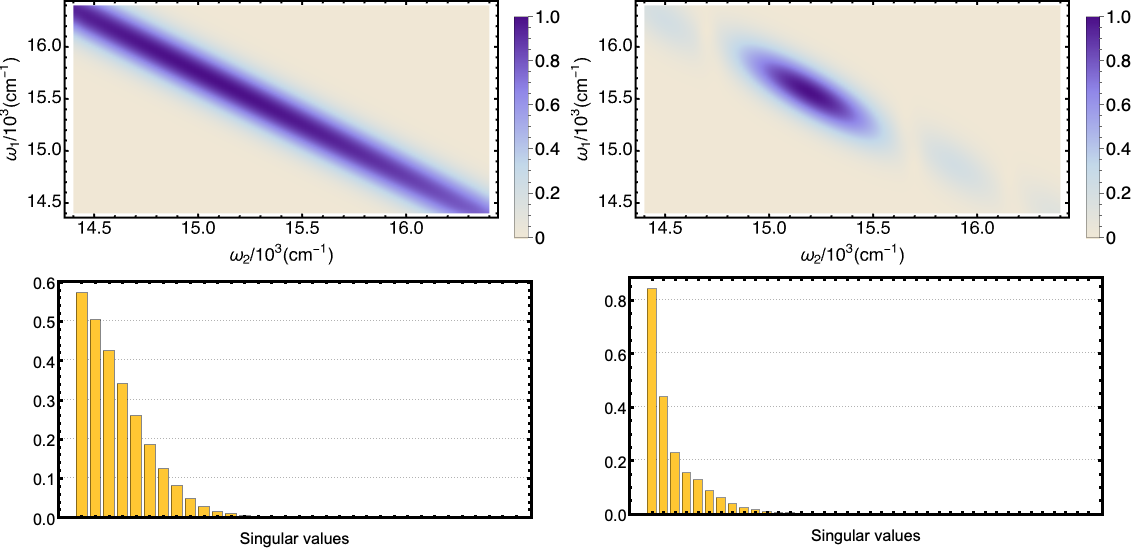}
\caption{The squared Joint Spectral Amplitude is presented for two distinct entanglement time parameters: $\tilde{T}_{\text{ent}} = 10$ fs (left column) and $60$ fs (right column). Two lower panels display the distribution of the $20$ largest normalized singular values.}
 \label{fig:entcorr}
\end{figure*}

For a better understanding of the correlation properties of the entangled photon pairs, in Fig.~\ref{fig:entcorr}, we plot the squared Joint Spectral Amplitude (JSA) and the singular values obtained via singular value decomposition of the kernel. An increase in the entanglement time parameter is inversely correlated with the number of statistically significant singular values. Consequently, the parametric freedom to select anti-correlated frequency pairs is diminished.

\bibliography{DQC}

\end{document}